\documentclass[aps,prl,twocolumn,superscriptaddress]{revtex4}
\usepackage{graphicx}
\usepackage{amssymb}

\begin{document}
\title{
%Usefulness of partial fluorescence yields 
%in x-ray absorption spectroscopy
Utility of the inverse partial fluorescence 
for electronic structure studies 
of battery materials}

\author{H.~Wadati}
\email{wadati@ap.t.u-tokyo.ac.jp}
\homepage{http://www.geocities.jp/qxbqd097/index2.htm}
\affiliation{Department of Applied Physics and Quantum-Phase Electronics 
Center (QPEC), University of Tokyo, Hongo, Tokyo 113-8656, Japan} 
\affiliation{Department of Physics and Astronomy, University of British Columbia, 
Vancouver, British Columbia V6T 1Z1, Canada}

\author{A. J. Achkar}
\affiliation{Department of Physics and Astronomy, 
University of Waterloo, Waterloo, Ontario N2L 3G1, Canada}

\author{D. G. Hawthorn}
\affiliation{Department of Physics and Astronomy, 
University of Waterloo, Waterloo, Ontario N2L 3G1, Canada} 

\author{T.~Z.~Regier}
\affiliation{Canadian Light Source, 
University of Saskatchewan, 
Saskatoon, Saskatchewan S7N 0X4, Canada}

\author{M. P. Singh} 
\affiliation{Regroupement Qu\'eb\'ecois sur les Mat\'eriaux de Pointe and 
D\'epartement de Physique, Universit\'e de Sherbrooke, Sherbrooke,
Qu\'ebec J1K 2R1, Canada}

\author{K. D. Truong} 
\affiliation{Regroupement Qu\'eb\'ecois sur les Mat\'eriaux de Pointe and 
D\'epartement de Physique, Universit\'e de Sherbrooke, Sherbrooke,
Qu\'ebec J1K 2R1, Canada}

\author{P. Fournier} 
\affiliation{Regroupement Qu\'eb\'ecois sur les Mat\'eriaux de Pointe and 
D\'epartement de Physique, Universit\'e de Sherbrooke, Sherbrooke,
Qu\'ebec J1K 2R1, Canada}

\author{G.~Chen}
%\affiliation{Department of Materials Science, 
%College of Materials Science and Engineering, 
%Jilin University, Changchun 130012, People's Republic of China} 
\affiliation{College of Physics/State Key Laboratory of Superhard
Materials, Jilin University, Changchun 130012, People's Republic of China }

\author{T.~Mizokawa}
\affiliation{Department of Complexity Science and Engineering, 
University of Tokyo, Kashiwa, Chiba 277-8561, Japan}

\author{G.~A.~Sawatzky}
\affiliation{Department of Physics and Astronomy, 
University of British Columbia, 
Vancouver, British Columbia V6T 1Z1, Canada}


\date{\today}
\begin{abstract}
X-ray absorption spectroscopy (XAS) 
is one of the most widely used experimental 
techniques to study the electronic and 
spatial structure of materials. 
Fluorescence yield mode is bulk-sensitive, 
but has several serious problems coming from 
saturation effects. In this study, we show 
the usefulness of partial fluorescence yields 
in addressing these problems. We discuss 
the different behaviors of 
La$_2$NiMnO$_6$ and LiMnO$_2$ at 
the Mn $2p$ absorption edges. 
The total fluorescence yield produces 
misleading spectra for LiMnO$_2$ due to the 
absence of high-$Z$ ($Z$: atomic number) elements. 
We conclude that the measurement of the inverse partial fluorescence 
yield is essential in studies of LiMnO$_2$, 
which is a hotly debated Li-ion battery material. 
%We conclude that the measurement of partial 
%fluorescence yield is essential in studies of LiMnO$_2$, 
%which is a hotly debated Li-ion battery material. 
\end{abstract}
\maketitle
%\section{Introduction}
X-ray absorption spectroscopy (XAS) is an 
experimental probe of the electronic and 
spatial structure of materials and 
is widely used in various fields of the natural 
sciences \cite{XASbook,Stohr,deGrootJES}. 
XAS is the measurement of photo-absorption 
by excitation of a core electron into the 
unoccupied states as a function of photon energy. 
There are three measurement modes for XAS, 
the transmission mode, 
the electron yield (EY) mode, and 
the fluorescence yield (FY) mode. 
In the transmission mode, the intensity of 
the x-rays is measured 
in front of and behind the sample and 
the ratio of the transmitted x-rays is determined. 
Transmission mode experiments are standard for hard x-rays, 
while for soft x-rays they are difficult to perform due to 
the strong interaction of soft x-rays with the sample. 
An alternative to the transmission mode experiment has been 
provided by measuring the decay products of the core holes 
which are created in the absorption process. 
The decay products include electrons, photons, 
and ions escaping from the surface of the sample. 
The yield modes can be classified into the EY mode and 
the FY mode. The total electron yield (TEY) is the 
most widely used detection technique. The energy of the 
outgoing electrons is not selected and simply all 
the escaping electrons are counted. The signal is 
dominated by secondary electrons which are created in 
the cascade process of the Auger decay electrons. 
The estimated probing depth of this technique 
using soft x-rays is $\sim40$ $\mbox{\AA}$. 
Besides the EY mode, the fluorescent decay 
of the core hole can be also used for XAS 
measurements. In the FY mode, the photons 
created in the fluorescent process have 
a mean free path of the same order of magnitude 
as the incident x-rays. Consequently, the FY 
mode has a large probing depth 
($>1000$ $\mbox{\AA}$ in the soft x-ray region), 
and is particularly suitable 
for studying bulk electronic structures. 
The FY measurements can be performed by either 
the total fluorescence yield (TFY) or 
the partial fluorescence yield (PFY). 
In the TFY mode photons with any energy are counted, 
whereas the PFY mode uses an energy-sensitive detector 
and selects photons within 
a particular range of energies. 
\begin{equation}
\begin{array}{rcl}
\mbox{TFY} & = & \mbox{PFY}_{\mathrm{Mn}}+
\displaystyle\sum_{X}\mbox{PFY}_{\mathrm{other}}\\
& = & 
C(E_f)\displaystyle\frac{\omega_{\mathrm{Mn}}(E_i,E_{\mathrm{Mn}})\mu_{\mathrm{Mn}}(E_i)}
{\mu_T(E_i)+\mu_T(E_{\mathrm{Mn}})\frac{\sin\alpha}{\sin\beta}}\\
& & +C(E_f)\sum_{X}\displaystyle\frac{\omega_X(E_i,E_X)\mu_X(E_i)}
{\mu_T(E_i)+\mu_T(E_X)\frac{\sin\alpha}{\sin\beta}},
\end{array}
\end{equation}
where
$$
\mu_T=\mu_{\mathrm{Mn}}+\sum_{X}\mu_X.
$$
Here, $\mu_X(E_i)$ is the contribution to the total attenuation 
coefficient from the excitation of core electron $X$, 
and $\omega_X(E_f)$ is the probability of fluorescence 
at energy $E_f$ resulting 
from electrons decaying to fill in the core hole left by $X$. 
$E_i$ is an incident photon energy. 
$C=\eta(E_f)\Omega/4\pi$ is a constant given by 
the detector efficiency $\eta(E_f)$ and the solid 
angle $\Omega$ of the detector. 
$\alpha$ and $\beta$ are the angles of 
incidence and emission, respectively, as measured 
from the sample surface. 

A significant limitation of FY mode XAS arises 
from saturation effects \cite{Eisebitt}, which 
distort peak intensities in the measured 
spectra and make FY non-linear with respect 
to the absorption coefficient. Recently, 
Achkar {\it et al.} introduced an alternative method 
to measure the absorption spectra which is 
bulk-sensitive and does not suffer 
from saturation effects \cite{Andrew}. 
They used an energy-sensitive detector 
and measured the normal (non-resonant) 
x-ray emission spectrum (NXES) from a 
different element and absorption edge 
than the one probed resonantly 
with incident photons. 
The inverse of this NXES PFY is 
termed as the 
inverse partial fluorescence yield
(IPFY), and is a function 
of the total x-ray absorption coefficient 
$\mu(E_i)$:
\begin{equation}
\mbox{IPFY}=A(\mu(E_i)+B), 
\end{equation}
where 
$A=4\pi/\eta(E_f)\Omega\omega_Y(E_f)\mu_Y(E_i)$, 
and $B=\mu(E_f)\sin \alpha/\sin \beta$. 
%$\alpha$ and $\beta$ are the angles of 
%incidence and emission, respectively, 
%as measured from the sample surface, $\eta(E_f)$ 
%is the quantum efficiency of the detector 
%at the emission energy, $\Omega$ is the detector solid angle,
%$\mu_Y(E_i)$ is the contribution to the total attenuation 
%coefficient from the excitation of core electron $Y$ (ex. O $1s$) 
%and $\omega_Y(E_i)$ is the probability of fluorescence at energy 
%$E_f$ resulting from electrons decaying 
%to fill in the core hole left by $Y$.
$Y$ denotes the core electron, for example, O $1s$. 
In Eqn.~(2), the constant $B$ is independent of $E_i$ and 
$A$ depends only weakly on $E_i$ over a narrow energy range 
and can be treated approximately as constant. 
Therefore, IPFY is proportional to the XAS $\mu(E_i)$ 
plus an offset proportional to $B$. 

In this paper, we show the different behaviors of 
La$_2$NiMnO$_6$ and LiMnO$_2$ at 
the Mn $2p$ absorption edges. One can 
see that the TFY makes peak structures in La$_2$NiMnO$_6$ 
and dip structures in LiMnO$_2$. 
This difference can be explained by the existence or absence 
of high-$Z$ ($Z$: atomic number) elements  such as La in this case. 
Then we show the usefulness of PFY and IPFY for 
the measurements of LiMnO$_2$. This technique should 
become an indispensable tool for the measurement 
of systems without high-$Z$ elements 
such as Li-ion battery materials. 

%\section{Experiment}
The thin film of La$_2$NiMnO$_6$ was fabricated 
on a SrTiO$_3$ substrate 
by the pulsed laser deposition technique, and 
the details were described in 
Refs.~\cite{MPSingh1, MPSingh2}. 
The powder samples of LiMnO$_2$ were 
synthesized by the procedure
described in Refs.~\cite{limno1, limno2}. 
The thickness of the thin films used 
for this study was 50 nm. 
XAS experiments were performed at 11ID-1 (SGM) of the
Canadian Light Source. 
The TEY spectra were measured using drain current. 
The TFY spectra were measured with a channel plate detector. 
The PFY spectra were measured using a silicon 
drift detector with an energy-resolution 
of $\sim120$ eV. 
The resolution power ($E/\Delta E$) was set to 5000.
All the spectra were measured at room temperature.

%\section{Results and Discussion}
Figure 1 shows 
the energy resolved x-ray emission spectra of 
La$_2$NiMnO$_6$ (a) and LiMnO$_2$ (b) 
(normalized to the incident photon intensity) 
as the incident photon energy is scanned 
through the Mn $2p$ absorption edges. 
Emission spectra for several incident photon 
energies are shown in Fig.~2, and 
we can see x-ray emission peaks corresponding to 
the following two transitions 
Mn $L_{\alpha,\beta}\sim$ 640 eV ($3d \rightarrow 2p$), and 
O $K_\alpha \sim 520$ eV 
%and O $K_{\alpha}$ 520 eV 
($2p \rightarrow 1s$). 
As expected, the Mn $L$ emission increases 
as the incident photon energy passes through 
the Mn $2p$ edges. 
In contrast, the O $K$ emission shows a small 
dip in the case of La$_2$NiMnO$_6$ (a), 
and a big dip in the case of LiMnO$_2$ (b). 
The La in La$_2$NiMnO$_6$ has a much larger absorption 
cross section that the other elements in the material 
and dominates the total attenuation coefficient. 
Therefore, even at the Mn $2p$ edges, 
the attenuation length of the incident photons does not change 
significantly and so the intensity of the O $K$ emission 
is not strongly reduced. 
When we measure the TFY from this sample, 
the absorption spectrum shows the expected positive structures, 
but they are slightly distorted by 
saturation effects, at the Mn $2p$ edges. 
The situation in LiMnO$_2$ is quite different. 
In this material, consisting of only low-$Z$ elements, 
the penetration depth decreases greatly when exciting 
across the Mn $2p$ edge.  This leads to a significant decrease 
in the O $K_{\alpha}$ emission that is 
larger in magnitude than the increase in Mn $L_{\alpha}$ emission. 
The resulting TFY spectrum exhibits dip structures 
as shown in Fig.~3. The dip structures are not apparent 
in the PFY. For PFY, a threshold photon energy of 600 eV 
can be used to separate the O $K$ and Mn $L$ emissions, 
as shown in Figs.~1 and 2. The PFY from the Mn $L$ 
emission ($>$ 600 eV) and the IPFY from the O $K$ emission ($<$ 600 eV) 
are shown in Fig.~3. 
\begin{figure}
\begin{center}
\includegraphics[width=7cm]{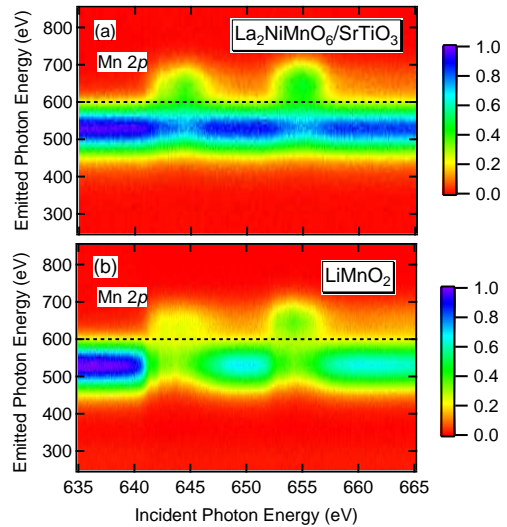}
\caption{(Color online): 
Energy-sensitive fluorescence yield from La$_2$NiMnO$_6$ and LiMnO$_2$. 
In panel (a) [(b)], normalized PFY of La$_2$NiMnO$_6$ [LiMnO$_2$] 
as the incident photon energy is scanned through the Mn $2p$ edges.}
\label{fig1}
\end{center}
\end{figure}

\begin{figure}
\begin{center}
\includegraphics[width=9cm]{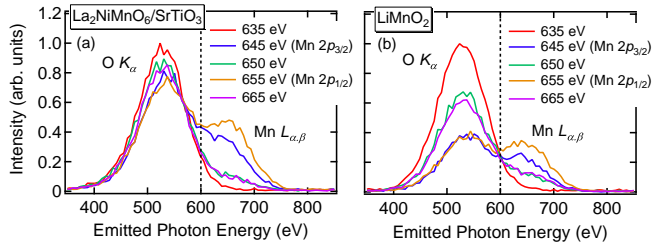}
\caption{(Color online): 
Emission spectra for several incident photon 
energies are shown for La$_2$NiMnO$_6$ (a) and 
LiMnO$_2$ (b). Emissions corresponding 
to the O $K_{\alpha}$ and 
Mn $L_{\alpha, \beta}$ transitions are observed.}
\label{fig2}
\end{center}
\end{figure}
%In the case of La$_2$NiMnO$_6$ (Fig.~3 (a)), 
%the TFY shows less distortion by saturation effects 
%than in La$_2$NiMnO$_6$. The Mn $L$ PFY displays 
%stronger saturation effects, and the IPFY from O $K$ 
%emission shows increased noise because the intensity of 
%O $K$ emission does not change significantly at this edge, 
%as shown in Fig.~2 (a). In the case of LiMnO$_2$ (Fig.~3 (b)), 
%the Mn $L$ PFY has already been shown in Ref.~[10], 
%and is less significantly distorted by saturation effects. 
%The IPFY derived from the O $K$ emission does not show 
%any saturation effects, 

In the case of La$_2$NiMnO$_6$ (Fig.~3 (a)), 
the TFY shows less distortion by saturation effects 
than in LiMnO$_2$. The Mn $L$ PFY displays 
stronger saturation effects, and the IPFY from O $K$ 
emission shows increased noise because the intensity 
of O $K$ emission does not change significantly 
at this edge, as shown in Fig.~2 (a). The IPFY 
of La$_2$NiMnO$_6$ also shows some distortion 
because the sample is only 50 nm thick. 
The IPFY method is based 
%on the assumption of total absorption conditions, 
%which are not being met in this case. 
on the condition of total absorption, 
which, in this case, is not met at all photon energies. 
For LiMnO$_2$ (Fig.~3 (b)), the Mn $L$ PFY has already 
been shown in Ref.~[10], and is less significantly 
distorted by saturation effects. The IPFY derived 
from the O $K$ emission does not show 
any saturation effects, unlike the PFY from Mn $L$ emission, 
consistent with the report in Ref.~[5]. 
This highlights that fact 
that TFY produces misleading spectra in many materials and 
that PFY is almost an indispensable tool to obtain 
bulk sensitive XAS spectra. 
Although reliable TEY spectra have also been measured 
on these materials, instances exist where surface contamination 
is significant or where the electronic states of the surface differ 
from the bulk \cite{LNO1,LNO2} 
and one must rely primarily on fluorescence detection. 
As demonstrated in this study, 
TFY is unsuitable for investigation of many Li-ion battery 
related materials, which is one of the most hotly debated 
topics in contemporary condensed-matter physics. 

\begin{figure}
\begin{center}
\includegraphics[width=8cm]{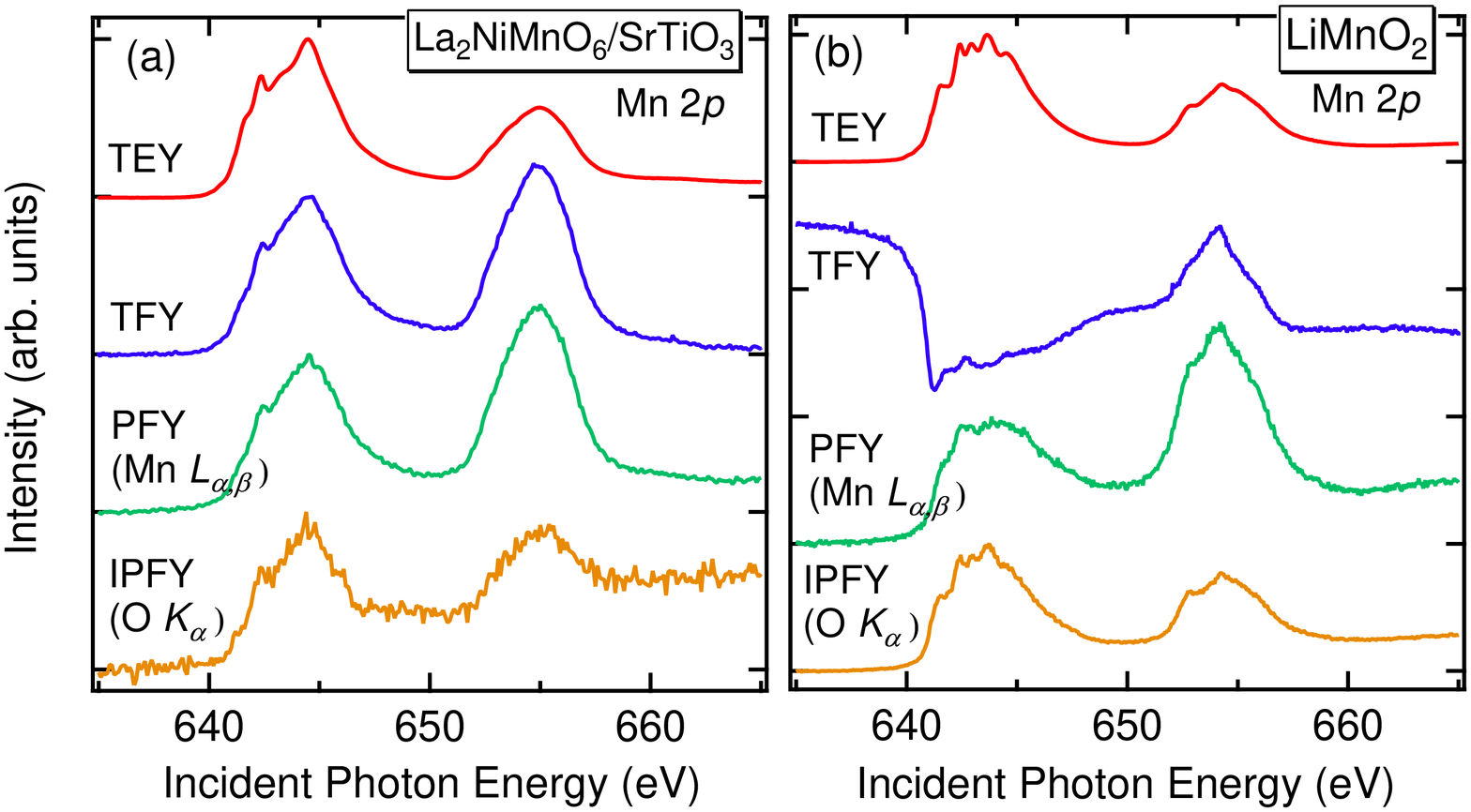}
\caption{(Color online): 
TEY, TFY, Mn $L$ edge PFY and 
O $K$ edge IPFY as a function of 
photon energy through the Mn $2p$ edges 
in the case of La$_2$NiMnO$_6$ (a) and 
LiMnO$_2$ (b).}
\label{fig3}
\end{center}
\end{figure}

%\section{conclusion}
In conclusion, we have demonstrated the indispensability 
of the IPFY technique for XAS studies of low-$Z$ battery 
related materials like LiMnO$_2$.  While TFY cannot 
be reliably applied to either LiMnO$_2$ or La$_2$NiMnO$_6$, 
there is significantly less distortion in the La$_2$NiMnO$_6$ 
which we attribute to the presence of the strongly absorbing La atoms 
which moderate the change in penetration depth when exciting across 
the Mn $L$ edge.  The PFY in both systems exhibits 
strong self-absorption effects. 
These effects are overcome by using the inverse 
of the O $K$ PFY (IPFY), 
which is a good representation of the Mn XAS in LiMnO$_2$. 
As such low-$Z$ systems are highly relevant battery materials, 
the IPFY technique will be a crucial tool for measuring 
accurate XAS spectra. 

%\section*{Acknowledgments}
This research was made possible with financial support 
from the Canadian funding organizations NSERC, CFI, and 
CIFAR. This research is granted by the Kao Foundation 
and also by the Japan Society for the Promotion of Science (JSPS) 
through the ``Funding Program for World-Leading Innovative R\&D 
on Science and Technology (FIRST Program),'' initiated by 
the Council for Science and Technology Policy (CSTP). 
\bibliography{LVO1tex}

\begin{thebibliography}{10}

\bibitem{XASbook}
F. de~Groot and A. Kotani, {\em Core Level Spectroscopy of Solids} (CRC Press,
  New York, 2008).

\bibitem{Stohr}
J. Stohr, {\em NEXAFS Spectroscopy} (Springer, New York, 1996).

\bibitem{deGrootJES}
F.~M.~F. de~Groot, J. Electron. Spectrosc. Relat. Phenom. {\bf 67},  529
  (1994).

\bibitem{Eisebitt}
S. Eisebitt, T. Boske, J.-E. Rubensson, and W. Eberhardt, Phys. Rev. B {\bf
  47},  14103  (1993).

\bibitem{Andrew}
A.~J. Achkar, T.~Z. Regier, H. Wadati, Y.-J. Kim, H. Zhang, and D.~G. Hawthorn,
  Phys. Rev. B {\bf 83},  081106  (2011).

\bibitem{MPSingh1}
M.~P. Singh, C. Grygiel, W.~C. Sheets, P. Boullay, M. Hervieu, W. Prellier, B.
  Mercey, C. Simon, and B. Raveau, Appl. Phys. Lett. {\bf 91},  012503  (2007).

\bibitem{MPSingh2}
M.~P. Singh, K.~D. Truong, S. Jandl, and P. Fournier, Phys. Rev. B {\bf 79},
  224421  (2009).

\bibitem{limno1}
Z.-F. Huang, F. Du, C.-Z. Wang, D.-P. Wang, and G. Chen, Phys. Rev. B {\bf 75},
   054411  (2007).

\bibitem{limno2}
F. Du, Z.-F. Huang, C.-Z. Wang, X. Meng, G. Chen, Y. Chen, and S.-H. Feng, J.
  Appl. Phys. {\bf 102},  113906  (2007).

\bibitem{WadatiAPL2010}
H. Wadati, D.~G. Hawthorn, T.~Z. Regier, G. Chen, T. Hitosugi, T. Mizokawa, A.
  Tanaka, and G.~A. Sawatzky, Appl. Phys. Lett. {\bf 97},  022106  (2010).

\bibitem{LNO1}
L.~A. Montoro, M. Abbate, E.~C. Almeida, and J.~M. Rosolen, Chem. Phys. Lett. {\bf 309}, 14 (1999).

\bibitem{LNO2}
Y. Koyama, T. Mizoguchi, H. Ikeno, and I. Tanaka, J. Phys. Chem. B {\bf 109}, 10749 (2005).

\end{thebibliography}

\end{document}